\documentstyle[aps]{revtex}
\begin{document}
\draft
\title{Gain without population inversion 
in V-type systems driven by a frequency-modulated field}
\author{Harshawardhan Wanare}
\address{Department of Physics, Indian Institute of Technology, 
Kanpur 208016, Uttar Pradesh, India}
\date{\today}
\maketitle
\begin{abstract}
We obtain gain of the probe field at multiple frequencies in a 
closed three-level V-type system using 
frequency modulated (FM) pump field.  
There is no associated population inversion 
among the atomic states of the probe transition.  We describe  
both the steady-state and transient
dynamics of this system.  Under suitable conditions, the 
system exhibits large gain simultaneously at series of frequencies 
far removed from resonance.  Moreover, the system can be tailored to 
exhibit multiple frequency regimes where the
probe experiences anomalous dispersion accompanied by negligible 
gain-absorption over a large bandwidth, 
a desirable feature for obtaining superluminal propagation of pulses
with negligible distortion. 
\end{abstract}
\pacs{42.50.Hz,42.50.Gy,32.80-t}

\section{Introduction}
In the last decade, a lot of remarkable effects of atomic coherence 
have been observed in various multilevel atoms \cite{review}. 
One of these effects is lasing without inversion (LWI) \cite{physrep}  
which has prospects in short-wavelength lasing.
Conventional lasers based on population inversion become impractical
in these wavelength regimes due to the $\omega^3$ dependence of the
Einstein A coefficient. Several experiments have provided
unequivocal evidence of LWI; starting from amplification without inversion
in transient \cite{awi1} then in the steady state regime \cite{awi2}, 
leading  ultimately to LWI \cite{lwie}.   
In the LWI schemes an external driving field
along a nearby transition generates atomic coherence which contributes 
to gain and alleviates the population inversion condition.
In this paper we show another significant contribution 
that becomes operative in presence of strong modulation and
results in gain at multiple frequencies. The conventional 
LWI based laser systems
suffer from a major limitation due to the small gain they exhibit in
comparison to the population inversion based laser systems.
However, this difficulty is largely overcome in this FM field driven 
system because under suitable conditions
one can obtain gain enhancement of ($\sim 10^3 \%$) with respect to
the conventional LWI schemes (monochromatic field driven systems).

The FM field interacting with a two-level system has been intensely 
studied earlier \cite{kaplan,hall,gsaold,usa1,nayak} and more recently 
\cite{usa2,ruyten,janowicz,sushi,hw}. 
The FM pump field produces a large number of sidebands \cite{janowicz}
which lead to periodic modulation of the absorption coefficient and the 
fluorescence signal \cite{nayak}. Analytical solutions have been obtained
for the weak modulation case in \cite{gsaold}.
The periodic FM field forms the basis of 
ultrasensitive absorption spectroscopy \cite{usa1,usa2} where the
resonant information is made to ride on the modulation 
frequency or its harmonics thus overcoming the predominantly
low frequency noise of lasers.
The FM field provides many 
independent parameters 
which can be sensitively controlled and thus result in
novel effects like the trapping of population \cite{hw}, in multilevel
systems suppression of a series of resonances occurs 
in the Autler-Townes spectra \cite{sfr}.  The trapping phenomenon due to
the periodic FM field \cite{hw1} or amplitude dependent phase modulation
\cite{warren} can be  exploited to achieve robust transfer of
population across multiple levels. 
The systems driven by amplitude modulated field have also been 
widely studied: the two-level system \cite{ampmod1} and the 
three-level system \cite{ampmod2} both exhibit multiple resonances resulting
from the modulation. We exploit the multiple resonances generated
by the FM field and
then tailor various parameters to obtain the desired gain features. 
To our knowledge 
studies have not been undertaken to obtain gain in multilevel
systems using FM fields, particularly in the strong modulation regime, 
we undertake this study in this paper.  

The organization of the paper is as follows: 
in section II we obtain the density
matrix equations that govern the dynamics of the V-type system pumped
by a FM field on one transition and is probed
by a monochromatic field on the neighboring transition.
We present the analysis of the results in section III, where we begin with
the description of the steady state response of the system and
go on to discuss the transient dynamics and finally
discuss the physical basis of the gain obtained. In the
steady state analysis we first deal with the off-resonant case
where the central frequency of the FM field is detuned from atomic
resonance followed by the on-resonance case.  
The contribution of atomic coherence between the
two excited states of the V-type system is known to be responsible for 
inversionless probe gain in systems pumped by a monochromatic field \cite{yzhu,gsa}. 
Here, we describe another contribution which comes 
into being  purely due to the modulation 
and plays a critical role in obtaining gain.
We also describe the Floquet analysis which
sheds light on the striking change that occurs in the probe spectrum
at specific values of the index of modulation. We utilize some
of these features to obtain anomalous dispersion for the probe field accompanied by
negligible absorption-gain in any desired frequency regime. This is an attractive feature 
for obtaining distortion free superluminal propagation \cite{wang}. 
Next, we describe the
transient dynamics and try to identify the dominant nonlinear processes
involved. The intricate dynamics of the time evolution is presented. 
We describe in detail the physical mechanism which is 
based on spontaneous emission assisted nonlinear optical process that
we believe dominates the gain process.
We present our conclusions in section IV.

\section{Model and Calculation}
 
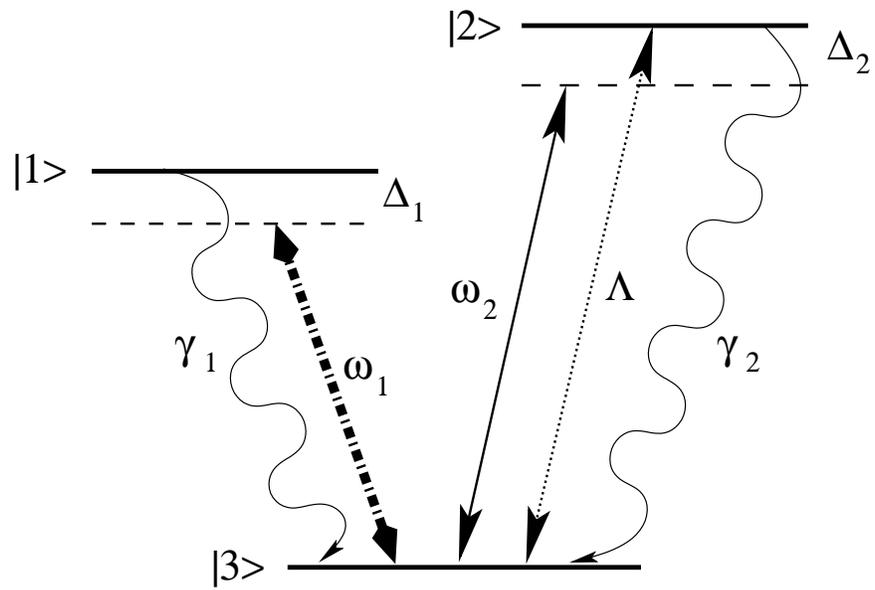
\begin{figure}
\caption{Schematic of the V-type system coupled to a frequency modulated
field on the $|1\rangle \leftrightarrow |3\rangle$ transition, the
central frequency $\omega_1$ of the FM field is shown with
a dash-dot line.  A monochromatic  probe field of frequency $\omega_2$
and an incoherent pump $\Lambda$ (dotted line)  couple the
$|2\rangle \rightarrow |3\rangle$ transition. The spontaneous emission decays from levels $|1\rangle$
and $|2\rangle$ to level $|3\rangle$ are represented by $\gamma_1$ and $\gamma_2$ 
(wavy line), respectively.
The detuning of the central frequency $\omega_1$ of the FM field
from the atomic resonance
is $\Delta_1$ and the probe detuning is $\Delta_2$.  }
\label{fig0}
\end{figure}

We consider a closed three-level V-type system (Fig. \ref{fig0}) wherein, one of the
transitions is coupled to a FM pump field 
and the other transition is coupled to a probe field as well 
as an incoherent (broadband)  pump. The field at the atom is given as
\begin{eqnarray}
{\bf E} = {\bf E}_1 e^{-i[ \omega_1 t + \Phi(t)]} 
&+&{\bf E}_2 e^{-i \omega_2 t} + c.c. ;\\ \nonumber
\Phi(t)=&M&\sin{(\Omega t)},
\end{eqnarray}
where ${\bf E}_1$ (${\bf E}_2$)  is the amplitude of the pump (probe) field, the FM field is
sinusoidally modulated about the 
central frequency $\omega_1$ and the modulation is characterized by
two independent parameters -
the frequency of modulation $\Omega$ and the index of modulation $M$.
The FM field couples the $|1\rangle\leftrightarrow |3\rangle$ 
transition and the monochromatic probe field at $\omega_2$ couples the
$|2\rangle\leftrightarrow |3\rangle$ transition. 

The total Hamiltonian of the system is 
\begin{equation}
H = \hbar \omega_{13}|1\rangle \langle 1| + \hbar \omega_{23}|2\rangle \langle 2| 
- {\bf d} \cdot {\bf E},
\end{equation}
where, ${\bf d }= {\bf d}_{13} |1\rangle \langle 3| + {\bf d}_{23} 
|2\rangle \langle 3| + c.c.$ The first two terms in the Hamiltonian correspond to
the unperturbed atomic system where the energies are measured
from the ground state $|3\rangle $, and the last term is the interaction
term in the dipole approximation. The semi-classical density matrix equation
is
\begin{eqnarray}
\nonumber
\frac{d \rho}{dt} = \frac{i}{\hbar}[H,\rho] 
-\gamma_1 (|1\rangle \langle 1| \rho
-2 \rho_{11}|3\rangle \langle 3| + \rho |1\rangle \langle 1|)
&-&(\gamma_2 +\Lambda)(|2\rangle \langle 2| \rho
-2 \rho_{22}|3\rangle \langle 3| + \rho |2\rangle \langle 2|)\\
&-&\Lambda (|3\rangle \langle 3| \rho
-2 \rho_{33}|2\rangle \langle 2| + \rho |3\rangle \langle 3|),
\label{master}
\end{eqnarray}
where,
$2\gamma_{1}(2\gamma_{2})$ is the rate of spontaneous emission from the level 
$|1\rangle(|2\rangle)$ to $|3\rangle$, $2\Lambda$ is the rate of
incoherent pumping
on the $|2\rangle\leftrightarrow |3\rangle$ transition.  

We transform the
equation of motion (\ref{master}) into a frame rotating with the instantaneous
frequency of the field by using the following relations
\begin{eqnarray}
\nonumber
\tilde \rho_{ii}& =& \rho_{ii},~ i=1,2,3, \\
\nonumber
\tilde \rho_{13}& =& \rho_{13} e^{i[\omega_1 t + \Phi(t)]},\\
\nonumber
\tilde \rho_{23}& =& \rho_{23} e^{i\omega_2 t },\\
\nonumber
\tilde \rho_{12}& =& \rho_{12} e^{i[(\omega_1-\omega_2 )t + \Phi(t)]},
\end{eqnarray} 
and undertake the rotating-wave approximation 
by neglecting the counter-rotating terms at nearly twice the
optical frequency, like $e^{\pm 2i[\omega_1 t + \Phi(t)]} $ and $
e^{\pm 2i\omega_2 t}$. This  approximation is valid if
$|d \Phi(t)/dt| \ll \omega_1$ which is valid
for typical frequency modulation in the optical regime.
Due to the coupling to the FM field the slowly 
varying density matrix ($\tilde \rho$) equations involve 
time-dependent detuning factors which go as $d \Phi(t)/dt$. 

We intend to obtain {\em exact} solutions of the density matrix equation
for arbitrary strength of the fields and arbitrary values
of the index of modulation. For this purpose we use the Fourier
decomposition 
\begin{equation}
\label{decomp}
\tilde \rho_{ij} = \sum_{n=-\infty}^{\infty} \rho_{ij}^{(n)}e^{-in\Omega t};~~
i,j = 1,2,3,
\end{equation}
which involves integral multiples of the 
frequency of modulation $\Omega$. 

On substituting the expression (\ref{decomp}) in the equation of evolutions of the slowly varying
density matrix ($\tilde \rho_{ij}$) and equating various powers of $\Omega$, 
we obtain the following infinite set of equations where $n$ is an integer that
varies from $-\infty$ to $\infty$.  The equations that govern the dynamics of the system 
are
\begin{eqnarray}
\nonumber
\frac{d \rho_{11}^{(n)}}{dt} &=& -(2\gamma_1-in\Omega)\rho_{11}^{(n)}+
iG_1(\rho_{31}^{(n)} -\rho_{13}^{(n)}), 
\\ \nonumber
\frac{d \rho_{22}^{(n)}}{dt} &=& -(2\gamma_2+2\Lambda-in\Omega)\rho_{22}^{(n)}
+ 2\Lambda\rho_{33}^{(n)}+iG_2(\rho_{32}^{(n)}-\rho_{23}^{(n)}),
\\ \nonumber
\frac{d \rho_{33}^{(n)}}{dt} &=& -(2\Lambda-in\Omega)\rho_{33}^{(n)}
+(2\Lambda+2\gamma_2)\rho_{22}^{(n)}+2\gamma_1\rho_{11}^{(n)}
-iG_1(\rho_{31}^{(n)}-\rho_{13}^{(n)})-iG_2(\rho_{32}^{(n)}-\rho_{23}^{(n)}),
\\ \nonumber
\frac{d \rho_{12}^{(n)}}{dt} &=& -(\gamma_1+\gamma_2+\Lambda-
i(\Delta_1-\Delta_2)
-in\Omega)\rho_{12}^{(n)}+iG_1\rho_{32}^{(n)}-iG_2\rho_{13}^{(n)}+
\frac{iM\Omega}{2}(\rho_{12}^{(n+1)}+\rho_{12}^{(n-1)}),
\\ \nonumber
\frac{d \rho_{13}^{(n)}}{dt} &=& -(\gamma_1+\Lambda-i\Delta_1-in\Omega)\rho_{13}^{(n)}+
iG_1(\rho_{33}^{(n)}-\rho_{11}^{(n)})-iG_2\rho_{12}^{(n)}+
\frac{iM\Omega}{2}(\rho_{13}^{(n+1)}+\rho_{13}^{(n-1)}),
\\ 
\frac{d \rho_{23}^{(n)}}{dt} &=& -(\gamma_2+2\Lambda-i\Delta_2-in\Omega)\rho_{23}^{(n)}+
iG_2(\rho_{33}^{(n)}-\rho_{22}^{(n)})-iG_1\rho_{21}^{(n)},
\label{density}
\end{eqnarray}
where $\Delta_1 = \omega_{13}-\omega_1$ 
is the detuning of the central frequency of the FM field  from the atomic
resonance on the $|1\rangle\leftrightarrow |3\rangle$ transition, the probe field 
detuning is $\Delta_2 = \omega_{23}-\omega_2$ on the $|2\rangle\leftrightarrow |3\rangle$ transition. 
The strength of the atom-field coupling is given by the Rabi frequency 
$2G_i = 2{\bf d}_{i3}\cdot{\bf E}_i/\hbar$
on the $|i\rangle\leftrightarrow |3\rangle$ transition, for $i=1$ and $2$ 
denoting the pump and probe Rabi frequencies, respectively.
The six equations in Eqs. (\ref{density}) and three
equations of motion for $\rho_{21}^{(n)}$, $\rho_{31}^{(n)}$
and $ \rho_{32}^{(n)}$ form a set of nine equations for each $n$. 
We note that the set of equations for  $n$ are coupled to
set for $n \pm 1$. The closure of the above system requires that 
$\rho_{11}^{(n)}+\rho_{22}^{(n)}+\rho_{33}^{(n)}=\delta_{n,0}$. 

We obtain exact non-perturbative solutions of the above
equations in both the  transient as well as steady-state regimes.
The transient solutions
are obtained by taking a large set of closed $(2N+1)\times 9$ 
first order coupled differential equations
where the harmonic index $n$ varies from $-N$ to $N$, 
and numerically integrating them using fourth-order Runge-Kutta routine. 
Needless to say, we have checked the convergence of the solutions by increasing 
$N$ and decreasing the step-size of integration. 
We obtain the steady-state solutions by setting the left hand side time derivatives
to zero in Eqs. (\ref{density}) and obtain tri-diagonal recurrence relations
which are solved using the infinite continued fraction 
technique \cite{risken,nayak}.

\section{Results and Discussion}
\subsection{Steady-state response}
We present the steady-state response of the system described
in section II. We consider two different cases, first one is the off-resonant case wherein  the 
central frequency of the FM field is detuned from  the atomic transition
($\Delta_1 \neq 0$), and the second case where the central frequency of the FM field is 
on resonance with the atomic transition. The second case permits
a comparison with the usual system pumped by a monochromatic field instead of the 
FM field \cite{yzhu}. In all our calculations the relevant frequency/time
variables are appropriately  normalized 
with the spontaneous emission decay on the pump transition, namely  $\gamma_1$.

In presence of the FM field on $|1\rangle\leftrightarrow |3\rangle$ transition,
the probe field on the 
$|3\rangle\leftrightarrow |2\rangle$ transition exhibits two combs
of absorption peaks which are slightly displaced from each other.
Each comb has frequencies that are separated by $\sim \pm \Omega$ 
and the relative shift between the combs depends on $M$, $\Delta_1$ and $G_1$.  
The overall spectrum  appears as a series of double peak structures displaced by 
$\sim \pm \Omega$.  The width of the resonances depend on the incoherent 
processes like the decays and the incoherent pump. Fig. \ref{fig1}a  shows a typical
absorption spectrum of the probe field 
modified by a strong FM field on the 
neighboring transition. In our notation, absorption of the probe field occurs when 
Im$(\rho_{32}^{(0)})<0$.
The two displaced comb like resonances result from
probing the two linearly independent set of Floquet states
resulting from the time periodic nature of the FM field-atom interaction.
We discuss these aspects in detail below. 

In presence of the incoherent pump $\Lambda$ one comb of frequencies
experiences gain ( Im$(\rho_{32}^{(0)})>0$ corresponds
to probe gain). The two 
set of combs continue to remain displaced and one of them is flipped 
over in the
opposite direction exhibiting gain as seen in 
Fig. \ref{fig1}b.

In terms of the density matrix equations the steady state  contributions to the probe 
gain/absorption can be resolved into three different terms, namely,
\begin{eqnarray}
\label{terms}
\nonumber
F_1 &=& \frac {i~G_2~ (\rho_{22}^{(0)}-\rho_{33}^{(0)})
~(\gamma_1 + \gamma_2 + \Lambda - i(\Delta_1 -\Delta_2))}
{(\gamma_1 + \gamma_2 + \Lambda - i(\Delta_1 -\Delta_2))
(\gamma_2 + 2\Lambda + i\Delta_2) + G_1^2}, \\
\nonumber
F_2 &=& \frac{G_2~G_1~ \rho_{13}^{(0)}}
{(\gamma_1 + \gamma_2 + \Lambda - i(\Delta_1 -\Delta_2))
(\gamma_2 + 2\Lambda + i\Delta_2) + G_1^2}, \\
\nonumber
F_3 &=& -\frac{M\Omega}{2} \times \frac{G_1~(\rho_{12}^{(1)}+\rho_{12}^{(-1)})}
{(\gamma_1 + \gamma_2 + \Lambda - i(\Delta_1 -\Delta_2))
(\gamma_2 + 2\Lambda + i\Delta_2) + G_1^2}, \\
\rho_{32}^{(0)}~& =&~ F_1~ +~ F_2~ +~ F_3.
\end{eqnarray}
The first term $F_1$ contains the contribution of the population difference
between the energy levels $|2\rangle$ and $|3\rangle$ (when $M=0$ it
corresponds to the conventional
population inversion term), the next two terms
correspond to the coherence term in the conventional modulation free ($M=0$) 
scheme \cite{yzhu,gsa}. Note that all the three components have the same
denominator.  The second term $F_2$ provides the probe response to the
dynamics at the central FM frequency $\omega_1$, or, 
to  the $n=0$ contribution
on the $|1\rangle\leftrightarrow |3\rangle$ transition.
The third term $F_3$, we call the modulation term,  
provides the contributions arising 
purely due to  modulation. The 
contributions from  the higher order terms, namely $|n| \geq 1$, are 
mainly channeled through
the $F_3$ term. Note that the $F_3$ term provides the
coupling of $n=0$ response to the $n=\pm 1$ term and this coupling
is proportional to $G_1M\Omega$,  moreover, the coupling is independent 
of the probe Rabi frequency $G_2$.
Similar expressions can be written for 
each $n$ and the coupling from higher order terms  
$n\pm 1$ always occurs through  $G_1M\Omega$ and is 
independent of $G_2$. No approximations
are made in writing out the various components in Eq. (\ref{terms}), all the
individual factors contained therein, like 
$\rho^{(0)}_{22}, \rho^{(0)}_{33}, \rho^{(0)}_{13}$ and
$\rho^{(\pm1)}_{12}$, are calculated exactly and thus contain 
all the higher order contributions.

The major contribution from the first two terms ($F_1$ and $F_2$) for a weak probe occur at
frequencies 
\begin{equation}
\label{dress}
\Delta_2 \rightarrow \frac{\Delta_1}{2}~ 
\pm ~\frac{1}{2}\sqrt{\Delta_1^2 + 4 G_1^2},
\end{equation}
which is similar to the usual monochromatic pump case. 
The third term $F_3$ provides the higher order contributions
which become significant in presence of strong modulation 
due to the coupling proportional to $G_1M\Omega$, as seen in Eq. (\ref{terms}). 
For the parameters chosen for Fig. \ref{fig1}, the contribution
from the $F_3$ term dwarfs the $F_1$ and $F_2$ contributions at all the
frequencies other than those given in Eq. (\ref{dress}),
see  Fig. \ref{fig1}c. Thus, in the strong modulation regime 
the modulation term dominates over the 
population inversion and the coherence terms of the usual monochromatic
pump case.

In order to understand the two comb of resonances one can
obtain the Floquet spectrum resulting from the FM field coupling 
the atomic transition.  Moreover,
it has been demonstrated that in presence of strong modulation 
and under certain conditions the Autler-Townes response exhibits
simultaneous suppression of semi-infinite number of resonances \cite{sfr}.
The conditions under which this occurs can also be obtained by 
analyzing the Floquet-spectrum.  Following
the approach developed by  Shirley \cite{floq} to obtain the Floquet states,
we consider exactly the effect of the strong FM field on the 
atomic transition and neglect the weak effects of 
dissipation and the weak probe field
because  $G_1, \Omega \gg \gamma_{1}, \gamma_{2}, \Lambda, G_2$.
In this limit it is advantageous to 
look at the dressed states $|\Psi_1 \rangle$ and  $|\Psi_3\rangle$ (dressed by the FM field) 
instead of the
bare atomic states $|1\rangle$ and $|3\rangle$. The Schr\"odinger equation
of the evolution of the corresponding  dressed state amplitudes 
$\psi_1$ and $\psi_3$ is
\begin{equation}
\label{schro}
i \frac{d}{dt}
\left(
\begin{array}{c}
\psi_1\\
\psi_3 
\end{array}
\right) = 
\left(
\begin{array}{cc}
\Delta_1 - M \Omega cos(\Omega t) & -G_1 \\
-G_1 & 0 
\end{array}
\right)
\left(
\begin{array}{c}
\psi_1\\
\psi_3 
\end{array}
\right).
\end{equation}
As the  Hamiltonian in Eq. (\ref{schro}) depends on time 
periodically, with a period $2\pi/\Omega$, one can obtain
two linearly independent solutions of Eq. (\ref{schro}) in the following form
\begin{equation}
\label{expand}
\psi^{\pm}_i(t) = e^{-i \lambda_{\pm}t} \sum_{n=-\infty}^{\infty}
\chi^{\pm,n}_i e^{-in\Omega t}~~ (i=1~{\rm and}~3),
\end{equation}
where the index $+$ and  $-$ distinguish the two solutions.
On substituting expression (\ref{expand}) into Eq. (\ref{schro}) 
we obtain an infinite
set of recursion relations for $\chi^{n}_i, ~i=1~{\rm and}~3 $
\begin{equation}
\label{recur}
\left(
\begin{array}{cc}
-n \Omega + \Delta_1 & -G_1 \\
-G_1 & -n \Omega 
\end{array}
\right)
\left(
\begin{array}{c}
\chi^n_1\\
\chi^n_3
\end{array}
\right) 
- \frac{M \Omega}{2}\left(
\begin{array}{cc}
1 & 0 \\
0 & 0
\end{array}
\right) \left[
\left(
\begin{array}{c}
\chi^{n+1}_1\\
\chi^{n+1}_3
\end{array}
\right) +
\left(
\begin{array}{c}
\chi^{n-1}_1\\
\chi^{n-1}_3
\end{array}
\right) 
\right] = \lambda
\left(
\begin{array}{c}
\chi^n_1\\
\chi^n_3
\end{array}
\right),    
\end{equation}   
wherein $\chi^{n}_i$ is coupled to its nearest neighbors
$\chi^{n\pm 1}_i$, for brevity we have dropped the $\pm$ superscript.
These recursion relations can be
written as an infinite matrix eigenvalue problem with the following
infinite dimensional Hamiltonian ($H_f$)
\begin{equation}
\label{infham}
\left(
\begin{array}{ccccccccccc}
..&:&:&:&:&:&\downarrow&:&:&:&..\\
..&-\frac{M \Omega}{2}&0&-(n-1)\Omega + \Delta_1&-G_1&-\frac{M \Omega}{2}
&0&0&0&0&..\\
..&0&0&-G_1&-(n-1)\Omega&0&0&0&0&0&..\\
..&0&0&-\frac{M \Omega}{2}&0&-n\Omega + \Delta_1&-G_1&-\frac{M \Omega}{2}
&0&0&..\\
\rightarrow&0&0&0&0&-G_1&-n\Omega&0&0&0&..\\
..&0&0&0&0&-\frac{M \Omega}{2}&0&-(n+1)\Omega + \Delta_1&-G_1&
-\frac{M \Omega}{2} &..\\
..&0&0&0&0&0&0&-G_1&-(n+1)\Omega&0&..\\
..&:&:&:&:&:&:&:&:&:&..\\
\end{array}
\right).
\end{equation}
Here, we have ordered the elements of the Floquet Hamiltonian $H_f$
such that $i$ runs over the states $1~{\rm and}~3$ before a change in the harmonic index $n$ 
which takes integer values from $-\infty$ to $\infty$.
The eigenvalues of the $H_f$ matrix are the quasienergies associated with
various levels of the dressed states.
The structure of the matrix (\ref{infham}) results in 
periodic eigenvalues, and
each dressed state has an  infinite ladder of quasienergy levels
separated by $\pm \Omega$. In other words the comb of resonances
correspond to probing the two linearly independent set of dressed states 
$+$ and $-$, whose energies given by $\lambda_{\pm} + 
n \Omega$, where $n$ takes integer values from $-\infty$ to $\infty$. 
Moreover, two levels belonging to 
different dressed states complement 
each other, i.e., $\lambda_+ - \lambda_- = \Delta_1$ modulo$(\Omega)$.

It has been shown in detail \cite{sfr} 
that when 
\begin{equation}
\label{lam0}
\lambda + m \Omega = 0
\end{equation}   
for some integer $m$
the determinant of the characteristic equation of the above eigenvalue
problem factorizes into determinants of two semi-infinite blocks 
\begin{equation} 
\label{character}
det[H_f - {\bf I} \lambda]_{\lambda=-m \Omega} 
~=~ - G_1^2 ~det[H_a - {\bf I} \lambda]~det[H_b - {\bf I} \lambda]~~=~0.
\end{equation}
The two semi-infinite blocks 
are such that the harmonic index $n$ in $H_{a}$ and $H_{b}$ 
runs from $-\infty$ to $m-1$ and $m+1$ to $\infty$, respectively.
The quasienergies are ambiguous within integer multiple of $\Omega$ because replacing
$\lambda$ by $\lambda + k \Omega$, where $k$ is an integer, leaves the 
Eq. (\ref{character}) unchanged.
The above factorization results from the structure of the Floquet 
Hamiltonian - if the integer $m$ happens to be $n$ in  matrix 
(\ref{infham}) then the row and column
marked by small horizontal and vertical arrows, respectively,  in the matrix 
Eq. (\ref{infham}) contains only one 
non-trivial element namely $-G_1$, corresponding to the $\chi^{m}_3$ term. 
Now if $\chi^{m}_3 = 0 $ then 
the half-infinite system of equations is closed 
resulting in  the above factorization, eq. (\ref{character}).
To determine bounded non-trivial solutions of the characteristic equation
(\ref{character}) one requires that one of the following conditions is
satisfied, either 
\begin{equation} 
\label{cond}
det[H_a + {\bf I}~ m\Omega ] =0 ~~~{\rm or}~~~
det[H_b + {\bf I}~ m\Omega ] =0. 
\end{equation}
The above conditions results in 
\begin{equation} 
\label{lam1}
\chi^{n}_1 =
\chi^{n}_3 = 0~~ {\rm for~ all}~~ n < m~~~~ {\rm or}~~~~ n > m 
\end{equation}  
depending on
which of the conditions in Eq. (\ref{cond}) is satisfied. 
The various harmonics of the dressed states $\chi^{n}_i$ and the
corresponding quasienergies $\lambda$ are determined numerically with 
the dimensions of the Floquet Hamiltonian matrix to be
$2(N+1)$ where N is a large positive integer and the harmonic index $n$
in eq. (\ref{infham}) varies from $-N$ to $N$.

We calculate 
the variation of the quasienergies as a function of
the modulation index $M$ for a fixed value of the pump field coupling strength,
its detuning and the modulation frequency. In Fig. \ref{fig2}a  the corresponding
parameters are $G_1=20 \gamma_1$,  $\Delta_1=20 \gamma_1$,  and  $\Omega = 30 \gamma_1$.  
Henceforth, we drop the harmonic superscript 
{\tiny $(0)$} from the density matrix variables as we
will deal with only the zeroth order atomic response at the frequencies 
$\omega_1$ and $\omega_2$ unless 
otherwise specified.
As seen in Fig. \ref{fig2}a, the  quasienergy level  plotted with a solid line
crosses the zero energy periodically at $M = 2.915,~4.06,~6.26,~7.35$ and so on.
At these values of $M$ the condition (\ref{lam0}) is satisfied. 
The probe spectrum exhibits only one set of absorption peaks 
separated by $\Omega$,
instead of the usual set of double peaks, as Eq. (\ref{lam1}) is satisfied. 
Fig. \ref{fig2}b highlights this feature for  
$M =2.915$ which is the first zero crossing of the
quasienergy level in Fig. \ref{fig2}a.
There is only single
peaked structure for  $\Delta_2 > 0$, whereas 
the spectra for $\Delta_2 < 0$ continues to exhibit the two set of peaks.
In Fig. \ref{fig2}b the probe response is shown for various values
of the incoherent pump $\Lambda/\gamma_1 = 0, 0.2, 0.4, 0.6, 0.8$ and $1.0$. 
In the absence of the incoherent pump 
the probe absorption spectra 
for $\Delta_2 < 0$ contains two comb of absorption dips,
whereas for $\Delta_2 > 0$ it contains a single comb of
absorption dips separated by $ \sim \Omega$. As the incoherent pump
rate is increased one set of absorption dips transform into  gain peaks.
The gain is maximum at about $\Lambda/\gamma_1 = 0.5$.
We have observed that for
parameters where the semi-infinite resonances are suppressed,
the surviving single comb of resonances almost always exhibit
gain in presence of the incoherent pump.

As the probe spectrum shows well separated gain peaks (for $\Delta_2 > 0$) 
the intermediate region accords the possibility of anomalous dispersion.
In these frequency regimes the probe beam can 
experience superluminal velocity due to anomalous dispersion \cite{wang}. 
Regions of anomalous dispersion ($d~{\rm Re}(\rho_{32})/d \omega_2 < 0$) 
accompanied by  a flat region of nearly zero  absorption 
(Im$(\rho_{32}) \approx 0$) is particularly desirable
in achieving superluminal propagation. 
In Fig. \ref{fig3}, the region between a set of gain peaks is
expanded and one can see the anomalous dispersion accompanied by a
flat region of negligible gain/absorption.  
The superluminal propagation in Ref. \cite{wang} is achieved 
in between Raman gain lines.
The anomalous region between the 
gain lines, in their system, is quite limited in terms of the available bandwidth
for the probe pulse. This is because the two fields responsible for the
Raman gain cannot be too far detuned from resonance.
The trade off between achieving sufficiently large gain so as to
obtain sharp change in dispersion in between the
gain lines, and having the gain lines separated far enough so as to obtain 
minimal gain at the line center severely restricts the bandwidth in their
system. 
Moreover, Doppler broadening further reduces the 
desirable anomalous region.  In most other schemes too one lacks the
control over the separation between closely spaced doublet 
exhibiting inversion, as the doublets arise from 
either hyperfine splitting or isotope shift \cite{chiao}. 
This limitation
is overcome in our system because the bandwidth depends on the
frequency of modulation and one can achieve a few hundred GHz modulation
frequency at optical frequencies. Another
advantage of this system is availability of multiple periodically separated gain peaks
so that one could choose to operate the probe far off resonance and thus the
possibility of decreasing the noise
arising from spontaneous emission \cite{new}. 
Hence, our system in principle accords
more flexibility in obtaining anomalous dispersion region of 
desired frequency bandwidth in a desired frequency regime. The 
possibility of tailoring  dispersion/absorption-gain
characteristics by controlling different parameters of the modulation
and the incoherent pump to overcome restrictions 
due to Doppler broadening and power broadening are added advantages of this
system.

We now discuss in detail the resonant case where the 
central frequency of the FM field 
is on resonance with the $|1\rangle \leftrightarrow |3\rangle$ transition and
contrast our results with the traditional monochromatic ($M=0$)  case. In the
monochromatic pump case,
gain arises purely from the
coherence between the excited states $|1\rangle$ and $|2\rangle$
which are not dipole connected, and, has been shown to occur without
inversion in any state basis \cite{yzhu}.  
We calculate the steady-state response of the system in presence of modulation
and Fig. \ref{fig4}a depicts the gain obtained as a function of 
the index of modulation
$M$. At $M=0$ one obtains the gain typical of the 
monochromatic case, whereas for strong modulation 
($M$ away from zero) one finds much larger gain.
The Im$(\rho_{32})$ at 
$M=10.17$ is $3.75 \times 10^{-5}$ in comparison to 
$3.05 \times 10^{-6}$ at $M=0$, a $\sim 10^3\%$ increment over the
monochromatic case.  It should be borne in mind that the dynamics
in presence of strong modulation is such that significant contributions
occur from the numerous side-band resonances leading to such large gain.

The absorption
peaks (sharp dips in Fig. \ref{fig4}a) occur at those values of $M$ 
for which $J_o(M)=0$, note that we have chosen  $\Omega >> \gamma_1$. This effect
is expected on physical grounds  as one can see from
the spectral content of the FM field, namely
\begin{equation}
\label{fmspec}
e^{i M \sin(\Omega t)} = \sum_{p=-\infty}^{+\infty} J_p(M) e^{i p \Omega t}, 
\end{equation} 
where $J_p(M)$ is the Bessel function of integer order $p$.
Whenever the resonant
central frequency is absent on the $|1\rangle \leftrightarrow |3\rangle$ 
transition the population in level $|3\rangle$ dominantly experiences absorption
on the $|3\rangle \leftrightarrow |2\rangle$ transition, and
the large value of $\Omega$ also ensures that spectral components 
at $\pm \Omega$ with the
weight of $J_{\pm1}(M)$ are far removed in frequency from the 
resonant line center. In between these peaks of absorption
one obtains gain.  The fact that one obtains gain only when $J_0(M) \neq 0$
points toward the spontaneous emission assisted nonlinear optical process
responsible for gain which will be discussed in detail at the
end of this section.

In terms of the density matrix the major contribution to the gain
is from the modulation term $F_3$ of Eq. (\ref{terms}) 
as seen in Fig. \ref{fig4}b. For $M=0$ the 
contribution from the inversion term $F_1$ 
is negative while the positive  coherence term $F_2$ completely
offsets this and is responsible for gain. 
The $F_3$ term is zero at $M=0$ and starts increasing as $M$ gets larger.
For strong modulation the 
$F_3$ term largely overshadows all other contributions and is responsible
for gain in between the zeros of $J_0(M)$.   
The smaller dips in the $F_3$ component, that occur in between the the zeros of the $J_0(M)$,
result from the next higher contribution arising from
$\omega_1 \pm \Omega$. These set of dips that occur when  $F_3$  is large
and positive are due to the zeros of the the Bessel $J_1(M)$.
The weights of the contribution at $\omega_1 \pm \Omega$ are
$J_{\pm 1}(M)$, these contributions are absent when $M$ takes values such that $J_1(M)=0$.
It should be noted that even though 
the contribution  from $\pm \Omega$ disappears at these values of $M$,  the contribution
from higher multiples of $\Omega$ continues to be significant 
resulting in large $F_3$ term and the resulting gain.
The steady state population distribution is shown in Fig. \ref{fig4}c.
At $M=0$, $\rho_{22}<\rho_{11}<\rho_{33}$ and this results in gain
without inversion in the bare state basis as well as 
no inversion for the Raman process. For $M=3.7$
where the gain is $1.6 \times 10^{-5}$ about $ \sim 400 \%$
more than the monochromatic case, the relative population distribution
is the same, i.e., $\rho_{22}<\rho_{11}<\rho_{33}$;
there is no inversion in the bare state basis as well as
no inversion for the stimulated Raman scattering (SRS) process. 
The two photon SRS inversion condition requires that
$\rho_{22} > \rho_{11}$. 
For very large values of $M$, not shown in figure,  
even though there is no inversion in the bare state basis 
($\rho_{22}<\rho_{33}$) there is Raman inversion for the
SRS process namely the $|2\rangle \rightarrow |3\rangle \rightarrow |1\rangle$ process,
because $\rho_{22}$ becomes larger than $\rho_{11}$.
Note that in all our discussions the feature of the gain being inversionless
is between the bare atomic states $|1\rangle$, $|2\rangle$ and $|3\rangle$. 

\subsection{Transient dynamics}
We discuss the time evolution of atomic polarization
and population distribution in the FM field pumped V-type system and how it
approaches the steady state values.
Due to the presence of a strong FM field the transient response 
of the system is very complex making it difficult to 
identify the physical mechanism responsible for gain.
We consider in detail one case in which one can identify the dominant 
process and is comparable to the 
monochromatic pump case. We will also
present a general case wherein the dynamics is more complex.

We choose the on-resonance case where $\Delta_1=\Delta_2=0$, the modulation 
frequency is large ($\Omega=80 \gamma_1$)  and the index of modulation
is chosen such that it is a zero of the first order Bessel function, i.e., 
$J_1(M)=0$ for $M=10.1734$. This particular choice is 
such that the FM field spectrum has the central 
peak at $\omega_1$ with weight $J_0(M)$, no peak
at $\omega_1\pm \Omega$ as $J_1(M)=0$, the next peak in frequency
is far removed and is at $\omega_1\pm 2\Omega$ with weight $J_2(M)$, and
other peaks at $\omega_1\pm n\Omega$ 
with $n>2$ and weight $J_n(M)$ are further removed from resonance. Choice
of large $\Omega$  and the appropriate $M$ ensures that the effects of 
$\omega_1 \pm n\Omega$ for $n \neq 0$ are minimal and the
population and atomic polarization dynamics appear to be somewhat  similar
to the monochromatic pump case. 

In Fig. \ref{fig5}a we show the evolution of the population in presence
of the incoherent pump with the initial
condition $\rho_{33}=1$ and all other $\rho_{ij}=0$ (i,~j=1-3). 
It is seen that at all times
$\rho_{22} < \rho_{33}$ and  $\rho_{22} < \rho_{11}$, 
therefore,
there is no population inversion in the bare state basis and there
is no Raman inversion for gain through SRS process, respectively.
Fig. \ref{fig5}b shows the  evolution of 
population in absence of the incoherent pump. The population in level 
$|2\rangle$ is negligible because the probability of the population to be excited
to level $|2\rangle$ is very small as $G_2 << G_1$ and $\gamma_i$. 
As expected the population oscillates between level $|1\rangle$ and
$|3\rangle$ and in steady state $\rho_{11} \approx \rho_{33} \approx 0.5$,  though
$\rho_{33}$ is slightly larger than $\rho_{11}$ due to the presence of
the FM field and the decays.  It is also
observed that in presence of the incoherent pump the
population and the atomic coherences reach their steady
state values earlier in time. 
Fig. \ref{fig5}c(d)  shows the evolution of
gain-absorption  for the probe field (FM pump field at $\omega_1$) 
with and without the incoherent pump. We note the following:
they both oscillate dominantly with the same frequency 
and nearly in phase with each other.
Both the probe and the pump field experience gain in the transient
regime with or without the incoherent pump. 
In presence of the
incoherent pump only the probe field experiences gain in the
steady state, whereas with $\Lambda=0$ both the fields 
experience absorption in the steady state. 

The probe field experiences gain 
just after $\rho_{33}$ reaches a minimum value favoring the stimulation process from 
$|2\rangle \rightarrow |3\rangle$ rather than the 
$|3\rangle \rightarrow |2\rangle$ transition. The probe
field experiences gain in the interval when $d\rho_{33}/dt>0$ 
and reaches its maximum value at a time
when $d\rho_{33}/dt$ is the steepest.
The FM pump field at the $\omega_1$ frequency experiences gain
just after $\rho_{11}$ reaches its maximum value. 
The contribution to the envelop of the pump
gain does not seem to come  from SRS process even though the
two-photon inversion with  $\rho_{11} > \rho_{22}$ is
present. The signature for an SRS process would be that the
absorption-gain response on the  $|1\rangle \leftrightarrow |3\rangle$ and
$|2\rangle \leftrightarrow |3\rangle$ transitions be out of phase. In other words,
emission on one transition accompanied by absorption on the other transition.
This is however not present in this case as the envelops oscillate in phase with each other
as seen in Figs. \ref{fig5}c and d.
We re-emphasize that the gain in the above parameter regime is 
qualitatively akin to the monochromatic case apart from the rapid but weak (small amplitude)
oscillations due to the FM field on the central pump frequency $\omega_1$ 
in Fig. \ref{fig5}d.

We  present in Fig. \ref{fig6} the temporal evolution 
of the atomic polarization  and the population 
dynamics  when the central frequency
of the FM field is detuned from the atomic resonance. 
The parameters chosen are identical to that in  Fig. \ref{fig1}
which corresponds to the steady state behavior. 
The population initially is in level $|3\rangle$.  The probe
frequency is tuned to the gain peak at $\Delta_2=46.8 \gamma_1$. We observe that
the population in levels $|1\rangle$ and $|3\rangle$ oscillate 
$\sim 180^o$ out of phase with respect to each other. The population in $|2\rangle$ 
is negligible in absence of the incoherent pump and
steadily increases in presence of the incoherent pump, see Figs. \ref{fig6}a-b.
At time  $\tau \gamma_1  \approx 3$ the population 
$\rho_{22}$ becomes larger than the
population $\rho_{11}$, however, 
$\rho_{22}$ remains smaller than $\rho_{33}$ at all times. Again there is  no population 
inversion between the levels $|2\rangle$ and $|3\rangle$.
After time $\tau \gamma_1 \approx 3$, unlike the resonant case shown in Fig. \ref{fig5},  
the population $\rho_{22} > \rho_{11}$ implying 
inversion for the SRS process for probe gain.

Both the probe and the pump fields experience gain in the
transient region with or without the incoherent pump.
It is clear that gain on the central FM frequency $\omega_1$ results
just after $\rho_{11}$ reaches its maximum value. 
Apart from the  high  frequency  oscillations
the envelop of the polarization on the central FM frequency $\omega_1$ closely 
follows the $\rho_{11}$ oscillations.  These high frequency
oscillations riding on the envelop result from higher order
contributions of field at $\omega_1 \pm n \Omega$ for $n>0$.  
There is no gain in the steady state  at the $\omega_1$ frequency,
whereas the probe field experiences gain in the steady state.

The frequency of oscillation of the probe field depends on its
generalized Rabi frequency and is found to be independent of the
incoherent decays and the incoherent pump. A closer look at the
probe oscillations in Fig. \ref{fig6}c shows that
periodically there are larger peaks of gain which invariably
follow just after $\rho_{33}$ reaches a minimum value, pointing to
similar process discussed above. Moreover, due to the
complexity of the time evolution  one cannot
rule out SRS process as some peaks of probe gain do
match with pump absorption dips and vice-versa.

\subsection{Physical mechanism for gain}
Although the quantitative results obtained above
from solving the density matrix equations provide
complete description of the system under consideration, 
it still does not provide insight into the underlying nonlinear
optical processes. Deeper understanding can be 
obtained by systematically summing up individual 
scattering processes and identifying the dominant contributions
to the gain.
We have observed that in the steady state
the probe gain occurs only if $\gamma_1 > \gamma_2$ \cite{comm}, implying
that spontaneous emission plays a crucial role in obtaining 
inversionless gain \cite{mossberg}.  
In recent times irreversible spontaneous emission assisted 
nonlinear optical processes have been suggested which contribute to
inversionless gain (in ladder-type systems by 
Sellin {\em et al} \cite{sellin} and in $\Lambda$-type systems by
Zhu \cite{yzhux}). Along these lines the physical picture
for inversionless gain in the steady state for the V-type system
would be the following. First a stimulated emission process from 
$|2\rangle \rightarrow |3\rangle$ of the probe
photon $\omega_2$, followed by absorption from 
$|3\rangle \rightarrow |1\rangle$ and finally spontaneous 
emission from $|1\rangle \rightarrow |3\rangle$ at the rate of $2\gamma_1$, 
see Fig. \ref{fig7}a.
Note that the probe experiences significant stimulated emission
only at certain frequencies due to the FM pump field, 
namely $\omega_2 \pm n\Omega$. These resonant features are absent
in the conventional LWI schemes. 
It should also be noted that in the above process the 
second step involving absorption from 
$|3\rangle \rightarrow |1\rangle$ depends critically on the 
availability of the $\omega_1$ photon (for $\Delta_1=0$). 
In its absence,
when the index of modulation $M$ is such that $J_0(M)=0$,  
this nonlinear process
is suppressed and the probe experiences only
absorption, see Fig. \ref{fig4}a. 
The competing process (Fig. \ref{fig7}b),
of stimulated emission from  $|1\rangle \rightarrow |3\rangle$
followed by absorption from $|3\rangle \rightarrow |2\rangle$
and spontaneous emission from $|2\rangle \rightarrow |3\rangle$
at $2\gamma_2 $ rate, is less preferred because of 
$\gamma_1 > \gamma_2$. Thus the former process
plays a  dominant role in inversionless probe gain. 

We have described above the nonlinear optical process responsible for gain 
when the central FM frequency $\omega_1$ is resonant with the atomic transition.
We have also looked at detuned cases and similar nonlinear process seems to
form the basis for obtaining gain. We describe briefly one such detuned case,
wherein $\Delta_1$ is finite and $\Omega$ is chosen to be equal to $\Delta_1$. 
It is clear that, in this case, the $\omega_1 + \Omega$ photon would 
be required in the second step of the nonlinear process (Fig. \ref{fig7}a) instead
of the $\omega_1$ photon. This is further supported by looking at the
polarization Im$(\rho_{32})$ as a function of $M$, and one observes sharp absorption
at those values of $M$ for which $J_1(M)=0$ (not shown here). This clearly shows that
in absence of the  $\omega_1 + \Omega$ photon there would be no 
spontaneous emission assisted gain process. Hence, the availability of the
intermediate photon which would take the population from level $|3\rangle$ 
to level $|1\rangle$
and the probe resonant with the periodic sideband response created
by the FM
field plays a crucial role in obtaining large gain.

It should be noted that the nonlinear optical process
discussed above is {\em assisted} by the
spontaneous emission and hence does not 
necessitate any kind of population inversion.  
The FM field plays a central role in creating
the rich periodic Floquet structure whose quasienergy levels 
are probed by the probe field. 
The on-resonance pump photon acts as a hub about which the
above nonlinear process occurs resulting in gain. 
We would also like to point out that if an amplitude modulated
field (with modulation frequency $\alpha$) is used instead of the FM field, 
then, due to the availability of only a pair of sideband photons 
($\omega_1 \pm \alpha$)
the gain from the above nonlinear process will be present only at 
a pair of frequencies. It is the availability of 
$\omega_1 \pm n\Omega$ photons
in the FM field that results in gain at a large number of 
sideband frequencies far removed from the central frequency.

\section{Conclusion}
In summary, we have presented an analysis of the  occurrence
of gain in a V-type system in presence of FM pump field. We have
analyzed both the steady state and the transient dynamics 
of the light amplification process. In the steady state analysis
we have tried to bring out the significant role played by the modulation
term over and above the coherence term which is dominant
in the conventional monochromatic pump schemes. We obtained the 
Floquet quasienergies that provide the explanation for the 
two comb of frequencies in the probe spectrum; furthermore the
quasienergy spectrum  was used to choose appropriate index of modulation $M$
such that the resulting probe spectrum contains a semi-infinite 
set of consecutive gain peaks far removed from resonance and could be exploited to 
obtain short-wavelength lasing.
In between the gain peaks 
anomalous dispersion is obtained accompanied by negligible 
gain-absorption, which could be utilized to obtain distortion free superluminal 
pulse propagation in any desired frequency regime and over a large bandwidth which 
can be increased by increasing the modulation frequency $\Omega$. 
The gain obtained in experiments based on 
LWI schemes is severely limited in its magnitude. The advantage
of gain  obtained using the FM field is the significant increment in its
magnitude $\sim 10^3 \%$ due to numerous sideband contributions,
and no population inversion
between the atomic states. 
It should be noted that the occurrence of gain at frequencies far 
removed from resonance
could also make this system attractive in terms of the 
possibility of reduced noise due to quantum fluctuations,
further investigations along these lines would be fruitful.
The study of transient dynamics showed that both the probe and the
pump fields experience transient gain. The probe gain is not due to
population inversion in the atomic state basis,
moreover, in certain regimes one could even rule out gain via 
SRS process and two-photon inversion.
We have described the physical process, based on
spontaneous emission assisted nonlinear optical process, which 
is responsible for inversionless gain. This effect
is further corroborated by the spectrum obtained
as a function of the index of modulation in Fig. \ref{fig4}a.


%
%
\newpage
\begin{center}
{\bf Figure Captions}
\end{center}

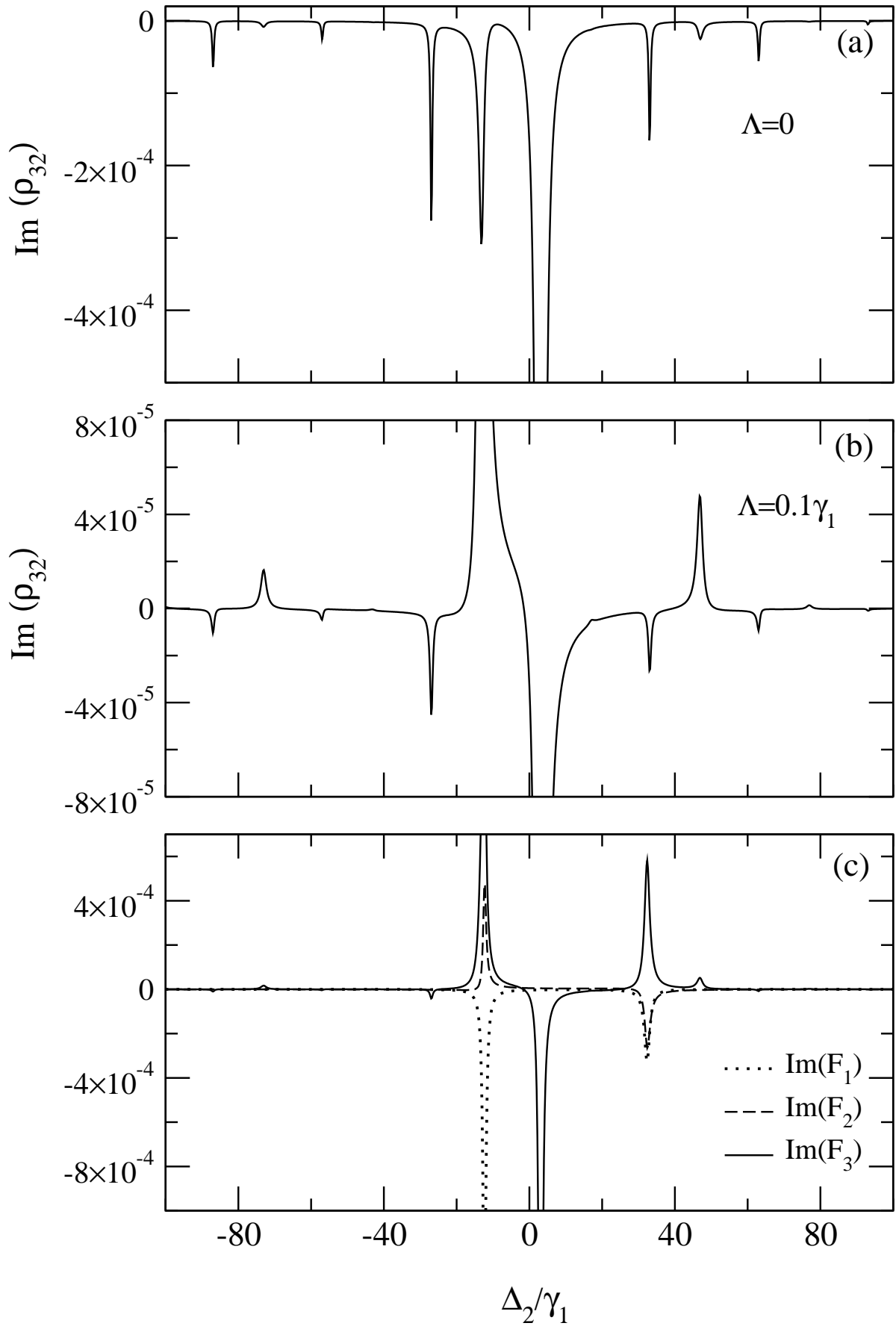
\begin{figure}
\caption{Steady state probe absorption-gain spectra - (a) Without the incoherent
pump the probe exhibits absorption as Im$(\rho_{32}) < 0$. (b) In presence of the incoherent
pump with $\Lambda = 0.1 \gamma_1$ alternate peaks exhibit gain. (c) The contribution
to $\rho_{32}$ from the terms $F_1$ and $F_2$ is significant only at frequencies given
by Eq. (\ref{dress}), whereas the contribution from the modulation term $F_3$ is the largest.
The other  relevant parameters are 
$M=5.2$, $\Omega = 30 \gamma_1$, $\Delta_1=20 \gamma_1$, $G_1 = 20 \gamma_1$,
$G_2=.01 \gamma_1$, and $\gamma_2 = 0.03 \gamma_1$. The scale on the graphs is chosen
to highlight the sideband structure and hence some of the maxima of the resonances
are not shown.  }
\label{fig1}
\end{figure}

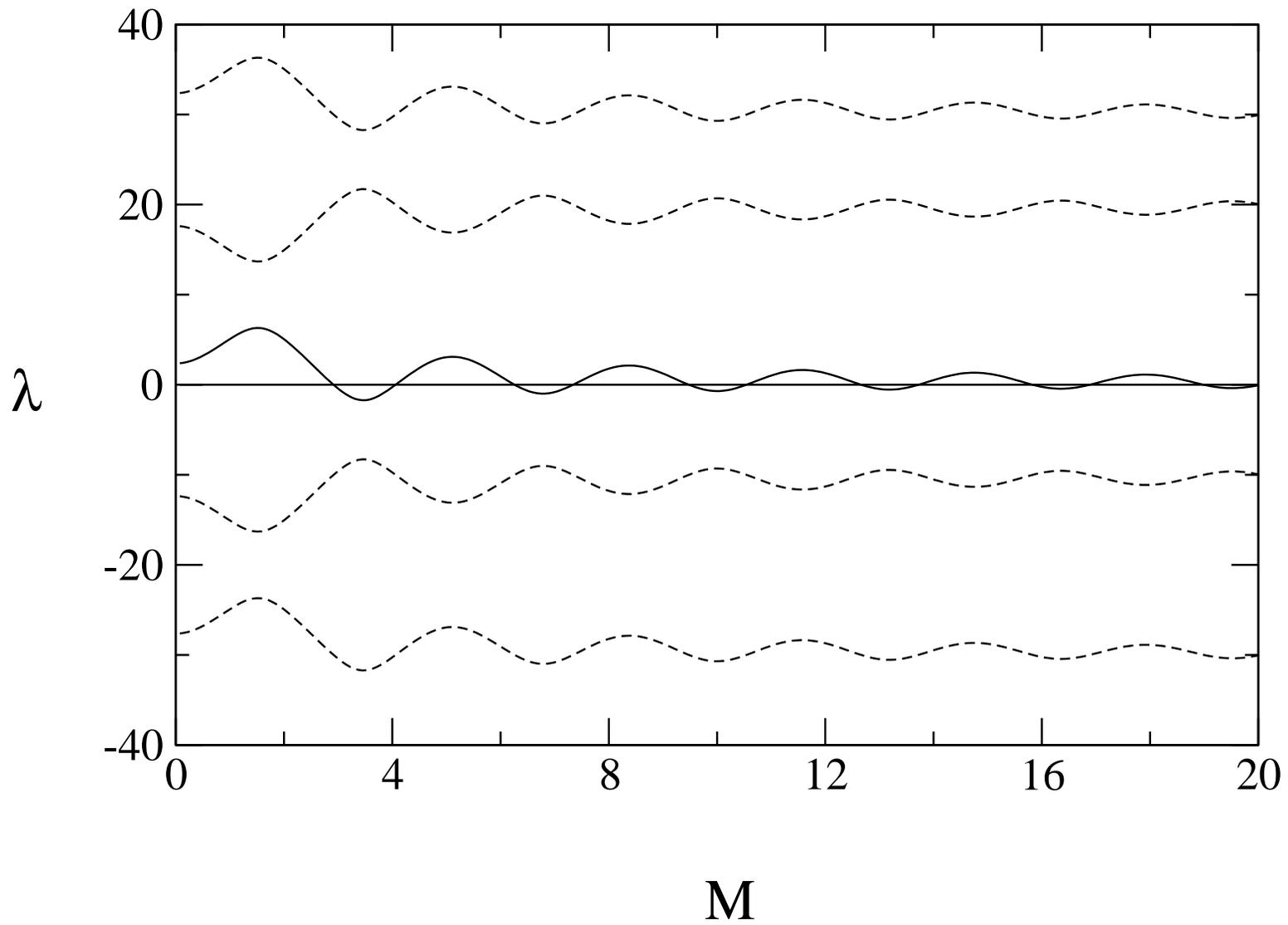
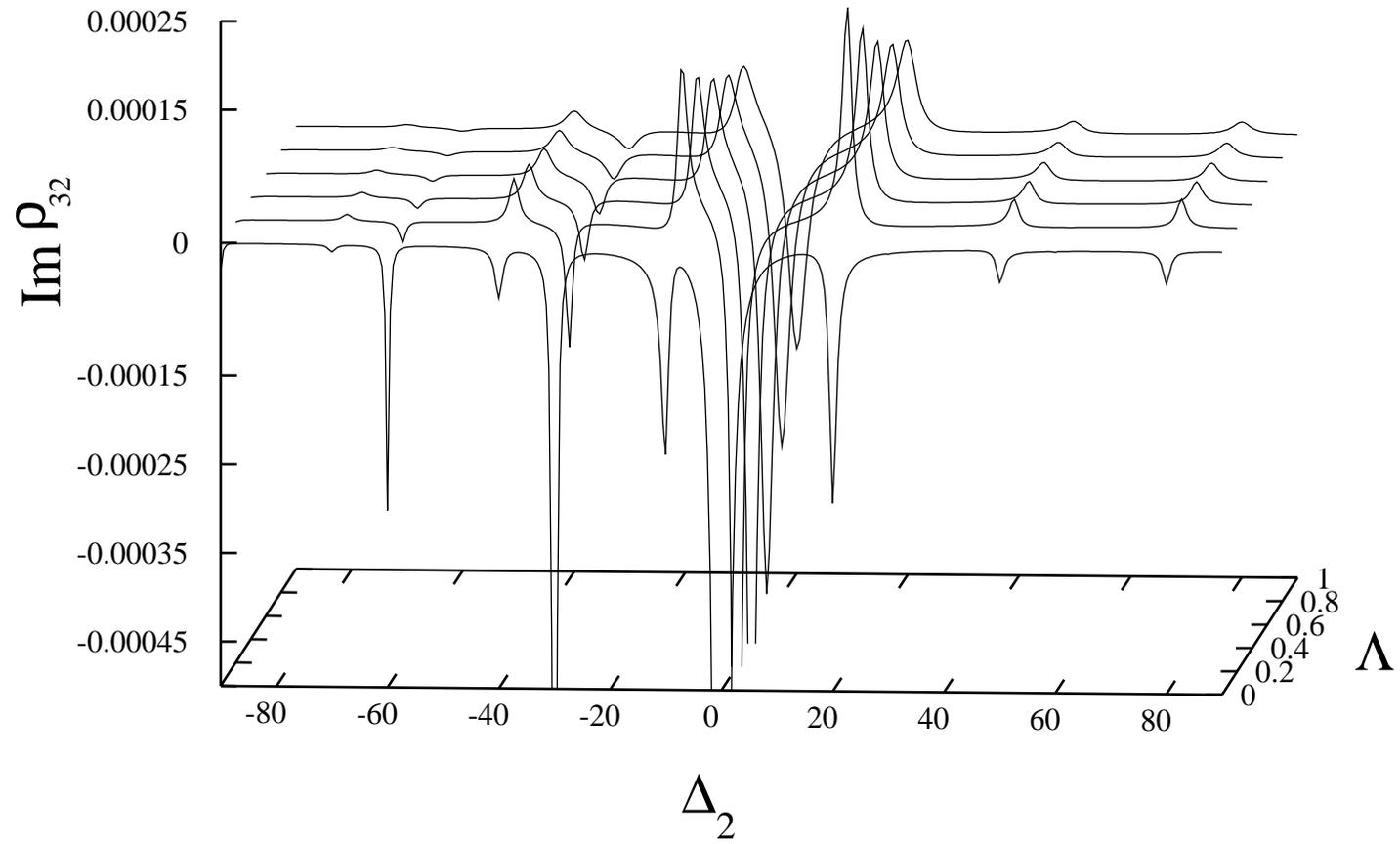
\begin{figure}
\caption{(a) Plot of a few quasienergies $\lambda$ vs M for a fixed value
of coupling field strength $G_1$ ($=20 \gamma_1$), detuning $\Delta_1$($ =20 
\gamma_1$) and the modulation frequency $\Omega$
($=30  \gamma_1$). The quasienergy level indicated by
a solid line periodically crosses the zero energy level. 
(b) Steady state probe response for various values of $\Lambda$ 
with the index of modulation $M = 2.915$ which is
the first zero of the quasienergy level shown in (a) by a solid line. 
The line plots are for incoherent
pump $\Lambda/\gamma_1~ = ~0, 0.2, 0.4, 0.6, 0.8$ and $1.0$. 
The other relevant parameters are
$G_2=.01 \gamma_1$, and $\gamma_2 = 0.03 \gamma_1$.
The scale on the graphs is chosen
to highlight the sideband structure and hence some of the 
maxima of the resonances
are not shown.  }
\label{fig2}
\end{figure}

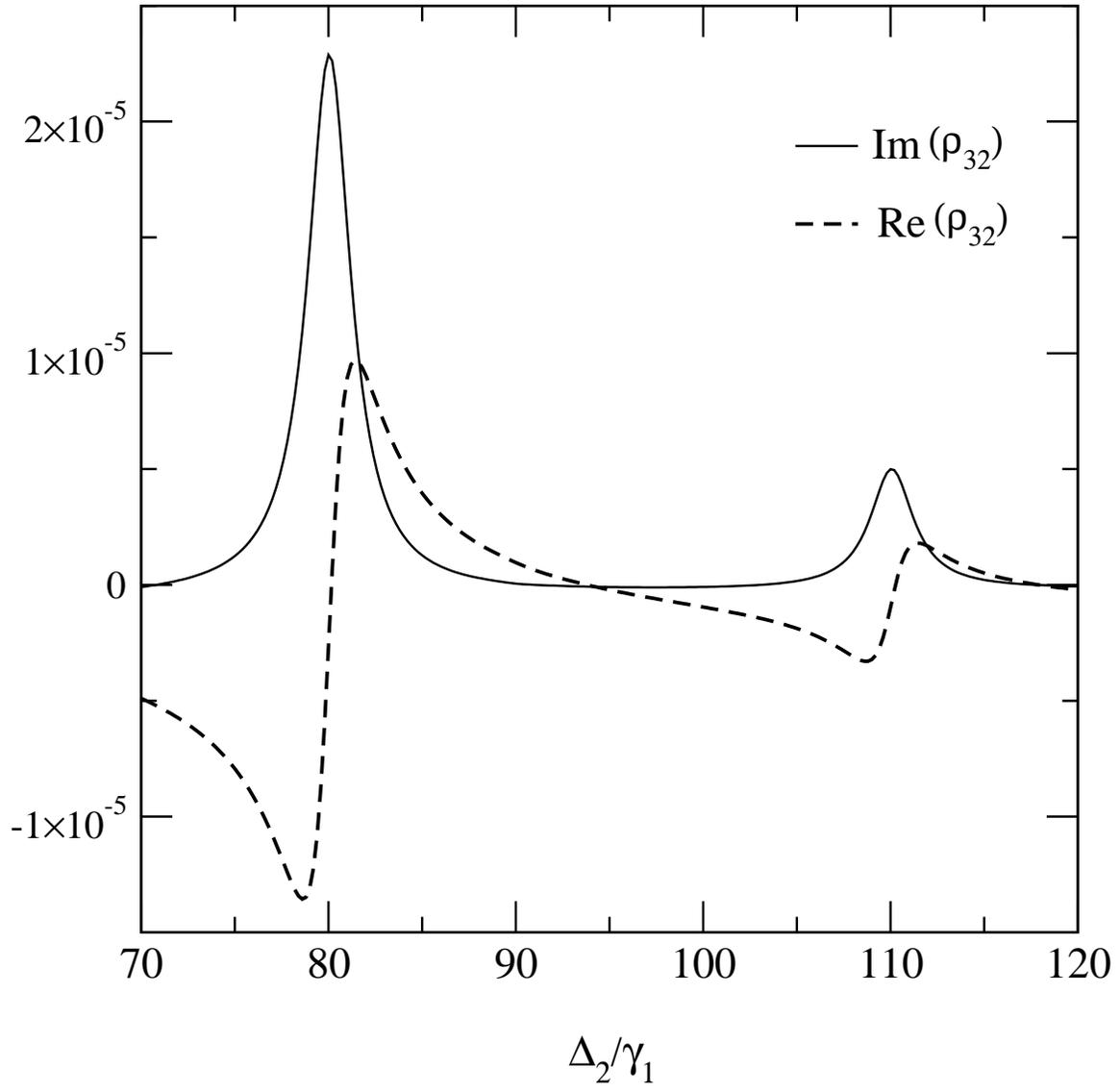
\begin{figure}
\caption{Enlarged plot of the atomic response in the region 
$\Delta_2/\gamma_1 = (70,120)$ with all the parameters same as Fig. \ref{fig2}
and $M=2.915$. In the region $\Delta_2/\gamma_1 = (90,100)$ the
probe field experiences anomalous dispersion [d Re($\rho_{32}$)/d$\omega_2$$< 0$] accompanied by 
negligible absorption/gain [Im($\rho_{32}$) $\approx 0$]. 
Such regions occur between each pair of gain peaks
for $\Delta_2 /\gamma_1 > 0$ in Fig. \ref{fig2}c. The slope of the Re$(\rho_{32})$
can be further increased by increasing the rate of the incoherent pump.}
\label{fig3}
\end{figure}

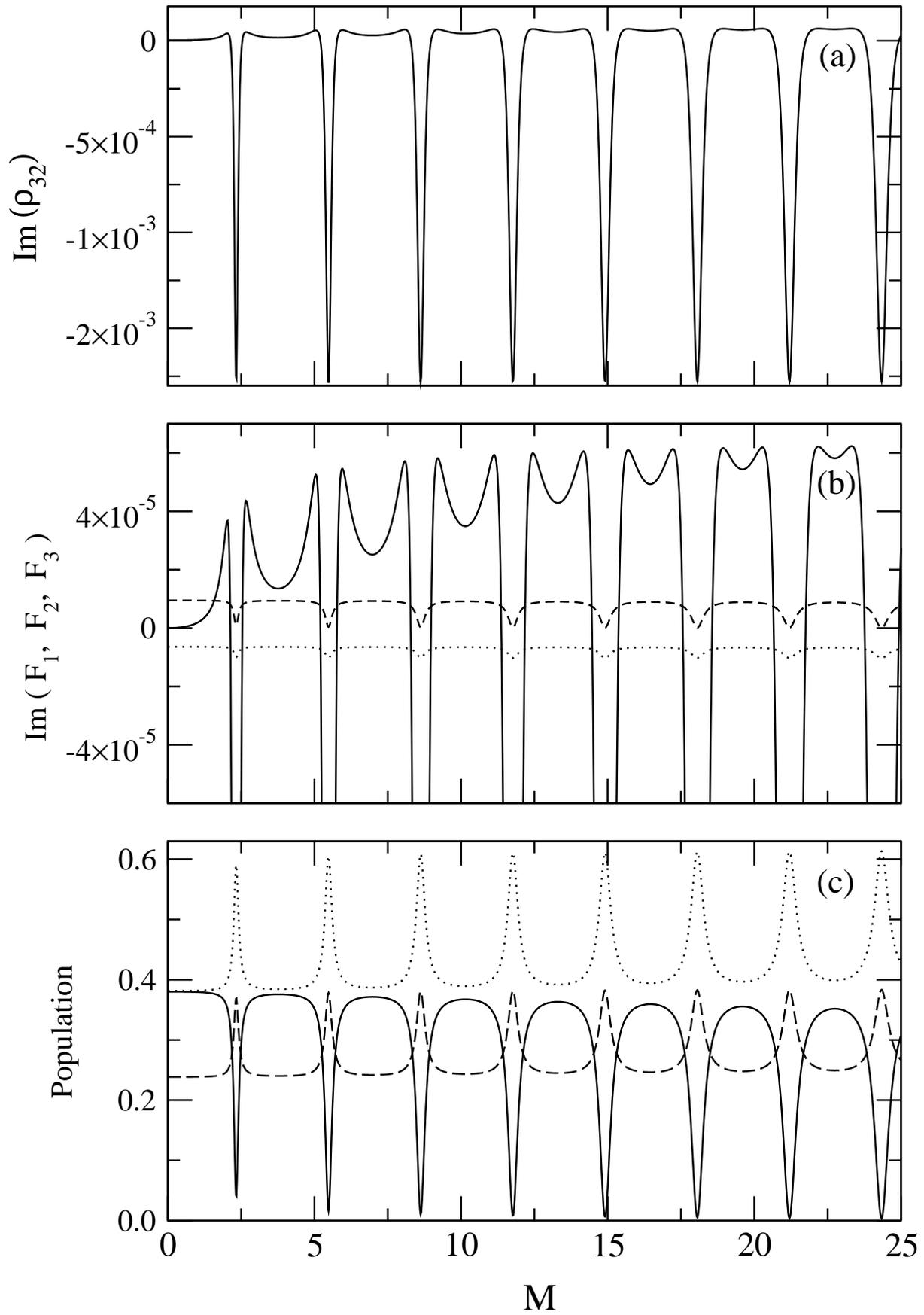
\begin{figure}
\caption{Plot of the resonant ($\Delta_1=\Delta_2=0$) steady state atomic response at the $|3\rangle \leftrightarrow |2\rangle$
transition as a function of the index of Modulation $M$. (a) Absorption-gain at the
probe transition, for $M=0$ Im($\rho_{32}$) = $+3.05 \times 10^{-6}$ which
is not clearly visible on the scale. (b) The individual  contributions $F_1, F_2$, and $ F_3$  denoted
by the dotted, dashed, and solid line, respectively. The $F_2$ contribution is responsible for
gain for weak modulation $0\leq M<1$; for strong modulation $F_3$ is the dominant term. 
To highlight the role played by the $F_3$ term in obtaining gain the large absorption minima
are not shown completely.  (c) The 
population in various atomic levels: the solid, dashed and the dotted  lines indicate $\rho_{11}$,
$\rho_{22}$, and $\rho_{33}$, respectively. The other relevant parameters are 
$\Omega = 80 \gamma_1$, $G_1 = 20 \gamma_1$,
$G_2=.01 \gamma_1$, $\gamma_2 = 0.3 \gamma_1$ and $\Lambda=0.5 \gamma_1$. }
\label{fig4}
\end{figure}

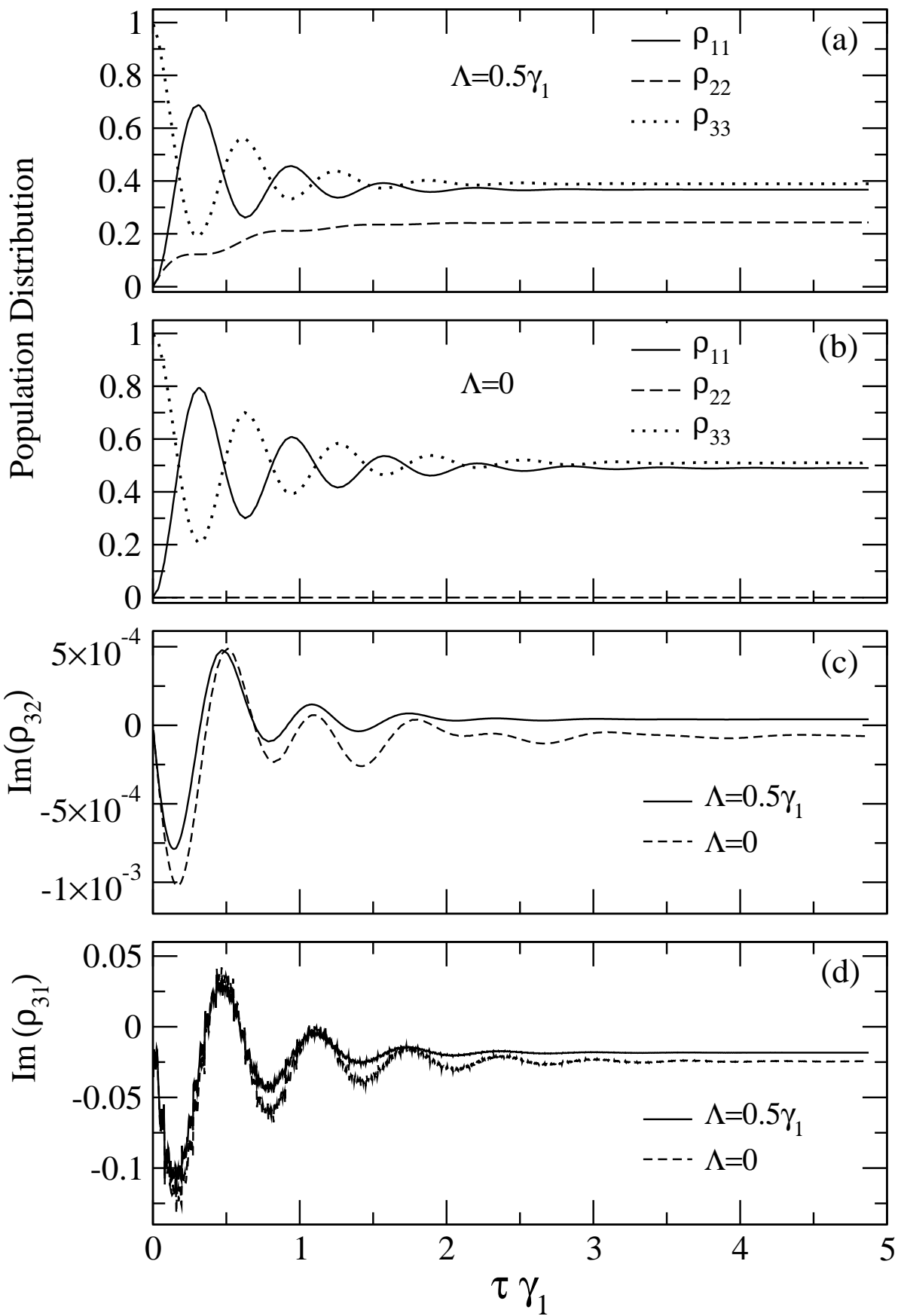
\begin{figure}
\caption{Calculated transient atomic response for the same parameters as in Fig. \ref{fig4} 
with the index of modulation $M=10.1734$ chosen to be a zero of the Bessel $J_1(M)$ function. (a) Population
distribution in presence of the incoherent pump ($\Lambda=0.5 \gamma_1$). (b) Population
distribution in absence of the incoherent pump ($\Lambda=0$). (c) Transient absorption-gain
response at the probe frequency  with and without the incoherent pump. (d)  Transient absorption-gain
response at the central frequency $\omega_1$ of the FM field  with and without the incoherent pump.
All the atomic variables are plotted against normalized time $\tau \gamma_1$.}
\label{fig5}
\end{figure}

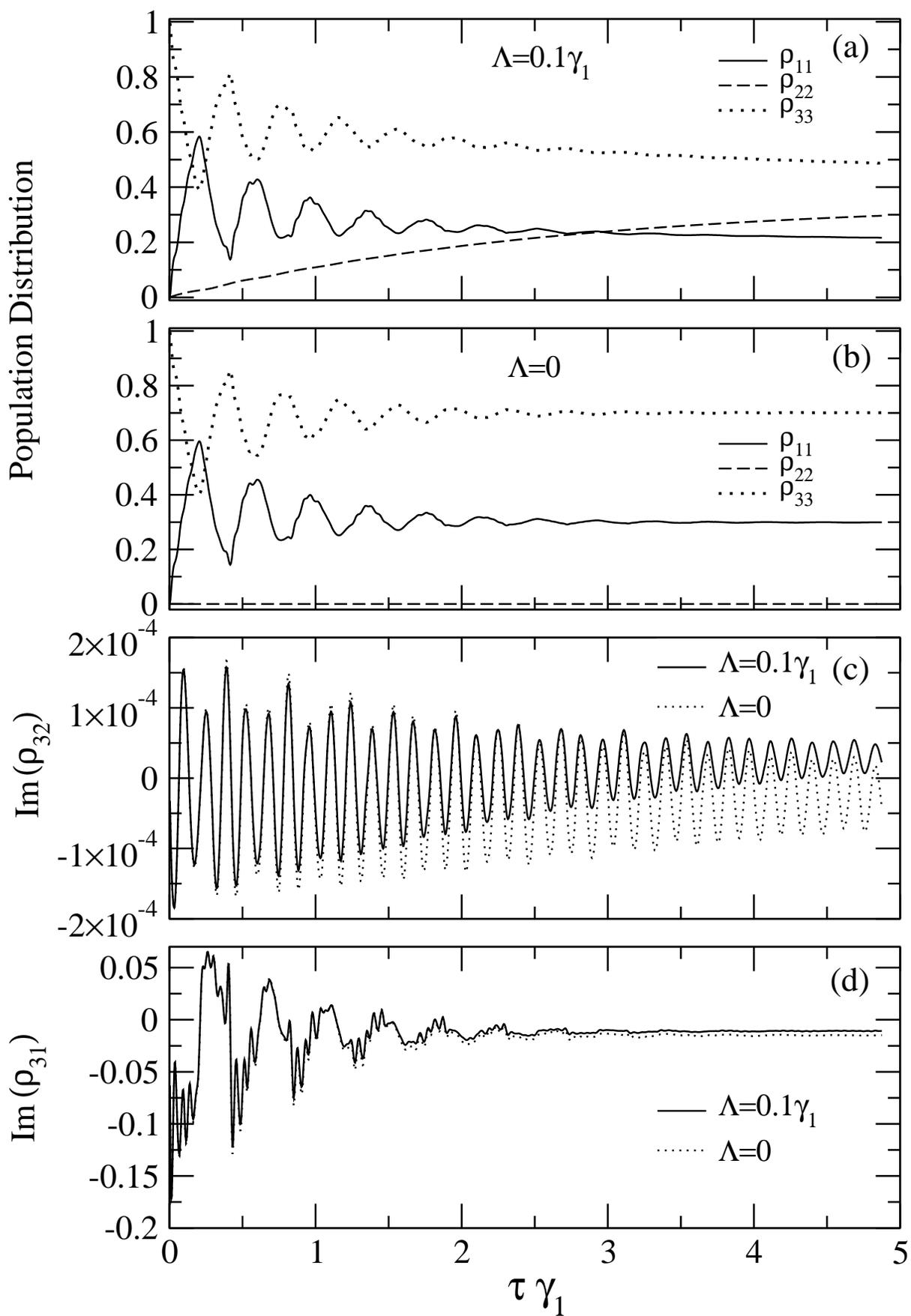
\begin{figure}
\caption{Calculated transient atomic response for the same parameters as in Fig. \ref{fig1}
at the probe detuning $\Delta_2=46.8 \gamma_1$. (a) Population
distribution in presence of the incoherent pump ($\Lambda=0.1 \gamma_1$). (b) Population
distribution in absence of the incoherent pump ($\Lambda=0$). (c) Transient absorption-gain
response at the probe frequency  with and without the incoherent pump. (d)  Transient absorption-gain
response at the central frequency $\omega_1$ of the FM field  with and without the incoherent pump.
All the atomic variables are plotted against normalized time $\tau \gamma_1$.}
\label{fig6}
\end{figure}

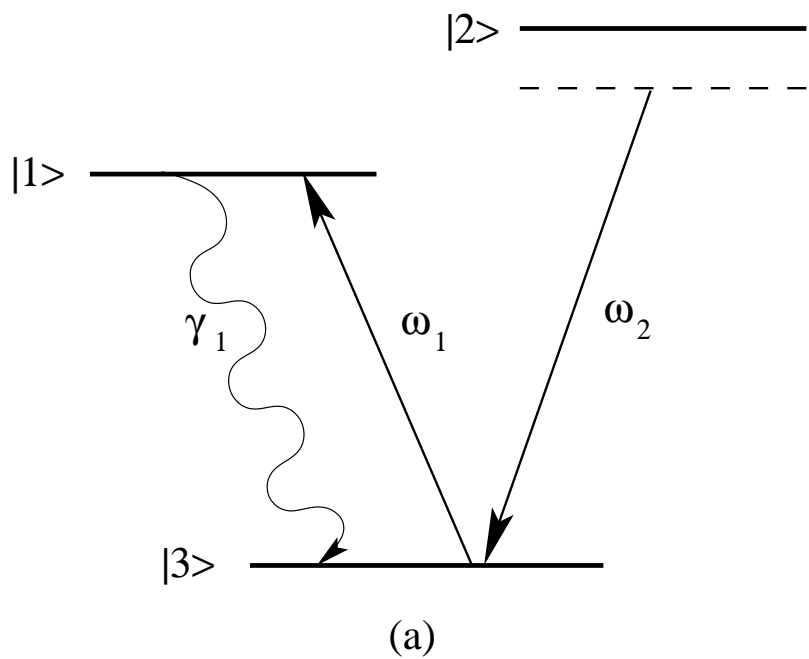
\begin{figure}
\caption{The physical mechanism, involving spontaneous emission assisted nonlinear optical 
processes, responsible for
inversionless gain. In steady state, process (a) prevails over process (b) because $\gamma_1 > \gamma_2$.
Note also that when $\Delta_1 = 0$, for process (a) to occur the photon $\omega_1$ has 
to be resonant with the 
$|1\rangle \leftrightarrow |3\rangle$ transition. In its absence
process (a) is suppressed resulting in probe absorption rather than probe gain.}
\label{fig7}
\end{figure}

%
%


\begin{references}
\bibitem{comm} The ratio $\gamma_2/\gamma_1$ is chosen to be much
smaller than unity in Figs. \ref{fig1} and \ref{fig2} to highlight the two comb
of resonances resulting from the two linearly independent set of solutions of the
Floquet Hamiltonian (\ref{infham}). The necessary condition for obtaining steady state
probe gain is  $\gamma_2<\gamma_1$.
\end{references}
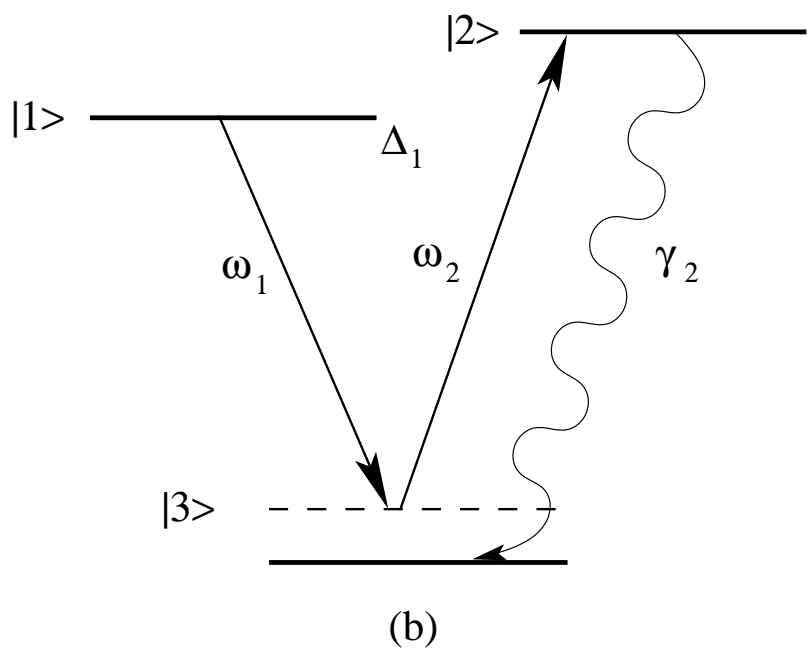
\end{document}